\documentclass[]{ceurart}

\usepackage{listings}
\usepackage{algorithm}
\usepackage{algorithmic}
\usepackage{booktabs}
\usepackage{tabularx}
\usepackage{subcaption}
\usepackage{amsmath,amssymb}
\usepackage{enumitem}
\usepackage{float}
\usepackage{placeins}

\begin{document}

\copyrightyear{2026}
\copyrightclause{Copyright for this paper by its authors.
  Use permitted under Creative Commons License Attribution 4.0
  International (CC BY 4.0).}

\conference{ITADATA2026: The 5$^{\text{th}}$ Italian Conference on Big Data and Data Science, November 10--12, 2026, Bari, Italy}

\title{Adaptive-Elbow Graph Traversal for Schema Retrieval in Text-to-SQL}

\author[1]{Fabio Picariello}[email=fabio.picariello@eng.it]
\cormark[1]
\author[1]{Salvatore Cipolla}[email=salvatore.cipolla@eng.it]
\cormark[1]

\author[1]{Susanna Bonura}[email=susanna.bonura@eng.it]
\author[1]{Vito Morreale}[email=vito.morreale@eng.it]

\address[1]{Engineering Ingegneria Informatica S.p.A., Piazzale dell’Agricoltura 24, 00144 Roma, Italy}
\cortext[1]{Corresponding author}

\begin{abstract}
LLM-based natural-language interfaces to relational databases are gated by an upstream
schema-retrieval step: the generator can only be as good as the slice of schema it is
shown. On a database with hundreds of tables, due to context size limitation, that slice must be selected and the selection
is two-sided---omit a needed table and the query becomes unrecoverable, include too many and
the irrelevant ones act as noise that measurably degrades the model's output. When an
agentic retriever that lets the model inspect the schema can be too costly or too slow, the
practical fallback is a lightweight local embedder that ranks tables by cosine similarity
and keeps a fixed top-$k$. This fallback has two weaknesses: a fixed budget is suboptimal
question by question, and embedding similarity overlooks foreign-key-linked tables that are
structurally necessary for joins. We present Adaptive-Elbow Graph Traversal (AEGT,
\textsc{ElbowHop} for short), a property-graph table retriever that addresses both: an
adaptive elbow cutoff replaces the fixed budget with a per-question one under a single
aggressiveness parameter, and bidirectional foreign-key traversal deterministically recovers
the bridge tables embedding similarity misses. We evaluate the retriever on three corpora of
increasing candidate-pool size (BIRD Mini-Dev queried per database, $3$--$13$ tables; its
eleven schemas merged, $74$; and a merged Spider-train graph, $546$) and isolate the two
mechanisms under a strict per-question budget-matched protocol. Across these settings the adaptive elbow emerges as the dominant mechanism---at matched
recall it roughly doubles precision over a fixed budget on the larger pools, the margin
widening as the pool grows---while the foreign-key traversal pays off on the join-critical
bridge tables dense retrieval cannot reach, in the single-hop tight-budget case. We characterize the method as a low-cost,
recall-robust first stage for schema retrieval on which a downstream selection or generation
step can build.
\end{abstract}


\maketitle

\section{Introduction}
\label{sec:intro}

Using LLMs to translate natural-language questions into SQL queries is a promising way to make enterprise data accessible without SQL fluency. However, their accuracy is upper-bounded by a step that takes place
before any query is generated: schema retrieval. A small schema can be shown to an LLM in
full; a typical enterprise database with hundreds of tables breaks this down---and not only by
exceeding the context window. Even when it fits, padding the prompt with irrelevant tables
measurably degrades accuracy: models are easily distracted by irrelevant
context~\cite{shi2023distracted}, use long inputs increasingly poorly as they
grow~\cite{liu2024lostmiddle}, and in RAG settings the noise from non-relevant passages can
dominate output quality~\cite{cuconasu2024noise}. The system must therefore first select a small
relevant subset, and the error is two-sided: miss the right tables and the correct query is
unrecoverable (a recall failure); bury them under wrong ones and the generator reasons worse than
it could (a precision failure). The upstream step thus caps the whole stack's accuracy.

One can let an LLM inspect the schema (reading descriptions, sampling values, reasoning
over joins) as the agentic stacks of MAC-SQL~\cite{wang2024macsql},
CHESS~\cite{talaei2024chess}, and CHASE-SQL~\cite{pourreza2024chase} do; but running such a stage
over the full table set of a large catalogue can be expensive, slow, and itself exposed to the
noise above. The regime we study puts a low-cost stage first: a local embedding model
ranks tables by cosine similarity and returns a small candidate set, so any downstream
stage like column selection, an LLM filter, or a full agentic selector---operates over a smaller subset. Our goal is not to replace the agentic stage but to make this cheap
prefilter as robust as possible: keep the gold tables (high recall) while shrinking the set to
cut noise.

The naive dense baseline in this regime has two chronic weaknesses. First, it returns a
fixed budget for all questions: a single $k$ is shipped to
production, yet simple questions get padded with distractor tables while complex ones lose
gold tables off the bottom of the list. Second, it ignores structure: tables that are
essential to the join graph but whose names and descriptions are semantically dissimilar
to the question---bridge tables such as \texttt{order\_items} for ``which products did
customer~$X$ buy?''---can be systematically missed. We attack the first weakness with an
\textbf{adaptive elbow} that sets a per-question budget under a single aggressiveness knob
$\alpha$, and the second with \textbf{bidirectional foreign-key traversal} over a property
graph of tables and columns.

Crucially, we do not assume these two mechanisms help equally, and we measure each in
isolation along an axis of growing difficulty: the size of the candidate pool the
retriever must search. We evaluate at the table level on three corpora---BIRD Mini-Dev
queried per database (a pool of $3$--$13$ tables), the same eleven schemas
merged into one $74$-table pool, and a merged Spider-train graph of $546$ tables.
Our contributions are:

\begin{enumerate}[leftmargin=*]
  \item \textbf{Adaptive elbow with aggressiveness scaling.} We extend the Kneedle
    knee-detection algorithm~\cite{satopaa2011kneedle} into a question-dependent retrieval
    budget controlled by a single $\alpha$ knob, replacing the fixed top-$k$ used by
    current text-to-SQL schema retrievers (Section~\ref{sec:elbow}).
  \item \textbf{Bidirectional foreign-key traversal.} We use a property graph of tables
    and columns to expand the seed set with bridge tables that embedding similarity cannot
    reach, in a deterministic step over a foreign-key graph loaded once from PostgreSQL and subsequently traversed in memory (Section~\ref{sec:traversal}).
  \item \textbf{An empirical map of when each mechanism pays off.} Under a strict
    per-question budget-matched protocol across three pool sizes---each reported gap
    attributable to one mechanism rather than to budget inflation---we show that the
    elbow's precision advantage at matched recall grows with the candidate pool, while the
    foreign-key traversal's recall contribution concentrates at tight seed budgets and a
    single hop (Sections~\ref{sec:protocol}--\ref{sec:res-combined}).
\end{enumerate}

We deliberately scope the present paper to the table stage---the object whose correctness gates
the join graph---and treat column selection and downstream SQL generation as separate stages,
the latter a direction we leave to future work.

\section{Background and Related Work}
\label{sec:related}

\paragraph{Schema retrieval and linking.}
Deciding which schema elements to put in front of the model is an old problem. Long before
LLMs, schema matching, surveyed by Rahm and Bernstein~\cite{rahm2001survey}, already
aligned a query with the right tables and columns by combining linguistic cues (their names)
with structural ones (how they connect in the schema). On the text-retrieval side, the
durable baseline for short, name-heavy strings such as table descriptions is Okapi
BM25~\cite{robertson1995okapi}; when a dense embedding ranker is available too, the two
ranked lists are typically merged into one by reciprocal rank fusion~\cite{cormack2009rrf}.
LLM-based text-to-SQL then made dense embedding similarity the default schema
retriever~\cite{talaei2024chess}. CRUSH4SQL~\cite{kothyari2023crush4sql} adds a twist:
instead of embedding the raw question, it has an LLM hallucinate the minimal schema
the answer would need and retrieves real tables against that imagined target.

A separate group of methods uses schema structure, but not to retrieve.
RAT-SQL~\cite{wang2020ratsql} and LGESQL~\cite{cao2021lgesql} fold the schema graph into the
encoder so the generator can see how tables and columns relate, and
RESDSQL~\cite{li2023resdsql} separates schema linking from generating the SQL skeleton; all
three assume the whole schema already fits in the model, so they never confront the
retrieval problem we do. The strongest agentic text-to-SQL systems make the same
assumption from the other end: MAC-SQL~\cite{wang2024macsql}, CHESS~\cite{talaei2024chess},
CHASE-SQL~\cite{pourreza2024chase}, and XiYan-SQL~\cite{gao2024xiyan} pour their engineering
into the downstream generator and selector and just reuse an off-the-shelf embedding
retriever upstream. That inherited retriever is exactly the first stage we set out to
improve, with their precision-oriented generation stage assumed to sit on top of it.

\paragraph{Graph-structured and adaptive RAG.}
Two lines of general retrieval-augmented generation~\cite{lewis2020rag} sit close to ours,
but each answers a different question than we do. The first asks where to retrieve
from, and answers with a graph: instead of treating a corpus as a flat list of passages, it
links the passages (or the entities in them) into a graph and searches over that. Microsoft
GraphRAG~\cite{edge2024graphrag} clusters related entities and retrieves one summary per
cluster, LightRAG~\cite{guo2024lightrag} builds a cheaper index and queries it at two
resolutions---individual entities and broad topics---and HippoRAG~\cite{gutierrez2024hipporag}
scores a passage by how a random walk over a knowledge graph reaches it. All of these graphs
are built over free text and the entities it mentions, not over a database's foreign keys;
StructGPT~\cite{jiang2023structgpt}, the one method here that does target structured data,
gives the LLM an interface to query the tables rather than a graph to retrieve them. None
uses the join structure a relational schema already carries---the structure our traversal
follows.

The second line asks whether to retrieve at all, rather than what:
Self-RAG~\cite{asai2023selfrag} trains the model to emit special tokens that, mid-generation,
call or skip a lookup; FLARE~\cite{jiang2023flare} triggers one whenever the sentence it is
about to write looks uncertain; and Adaptive-RAG~\cite{jeong2024adaptiverag} sends easy
questions down a light pipeline and hard ones down a heavy one. Neither line settles the
question we face---how many tables to keep once retrieval has fired, the cutoff
itself. That is a knee-detection problem, classically handled by the Kneedle
algorithm~\cite{satopaa2011kneedle}, which we extend with an aggressiveness parameter so the
budget--recall trade-off becomes a single tunable knob.

\paragraph{Positioning.}
Our retriever combines three ingredients in a configuration we have not seen elsewhere:
(i)~a property graph stores tables and columns as typed nodes with chunked embeddings, so
cosine retrieval and graph traversal share the same store; (ii)~the selection uses an
adaptive elbow with an explicit aggressiveness parameter that sets how much to retrieve per
question, a knob the adaptive-RAG
line~\cite{asai2023selfrag,jiang2023flare,jeong2024adaptiverag} does not offer, since it
decides only whether to retrieve, not how large the retrieved set should be; (iii)~bidirectional foreign-key traversal adds tables that are
structurally necessary for joins but semantically dissimilar to the query, a deterministic
complement to the LLM-hallucination bridge of CRUSH4SQL~\cite{kothyari2023crush4sql} and
the entity-community summarisation of Microsoft GraphRAG~\cite{edge2024graphrag}.

\section{Adaptive-Elbow Graph Traversal}
\label{sec:method}
\textbf{Adaptive-Elbow Graph Traversal} (AEGT; \textsc{ElbowHop}) maps each question to a table set $\mathcal{T}$ through adaptive semantic selection followed by bidirectional foreign-key traversal (Algorithm~\ref{alg:graphrag}). The resulting set feeds a downstream column-selection stage that retains primary and foreign keys and produces an \emph{M-Schema}-style representation~\cite{gao2024xiyan}; this stage is outside the present evaluation.

\subsection{Property Graph Store}
\label{sec:graphstore}

The schema knowledge graph is stored in PostgreSQL under a dedicated schema per tenant
through three relations: \texttt{nodes}(\textit{id, label, props JSONB}),
\texttt{edges}(\textit{id, src, dst, rel}), and \texttt{embeddings}(\textit{node\_id,
vector\_json JSONB}). Two node types are used by the retriever: \textbf{Table} (name,
description) and \textbf{Column} (name, datatype, description, is\_primary, is\_foreign),
linked by \texttt{HAS\_COLUMN} (Table $\to$ Column) and \texttt{REFERENCES}
(Table $\leftrightarrow$ Table, encoding foreign-key links and traversed bidirectionally).
As a reference scenario, consider the table-level path \[ \texttt{customers} \leftrightarrow \texttt{orders} \leftrightarrow \texttt{order\_items} \leftrightarrow \texttt{products}. \] Each relation is a Table node, its attributes are Column nodes connected through \texttt{HAS\_COLUMN}, and foreign keys induce the \texttt{REFERENCES} links shown above. For ``Which products did customer $X$ buy?'', semantic retrieval may favour \texttt{customers} and \texttt{products}, while traversal recovers the less salient intermediate tables.
\subsection{Embeddings and Cosine Similarity} Each node's text is embedded offline with \textbf{all-MiniLM-L6-v2} ($384$ dimensions, served locally). We also conducted preliminary experiments with the larger \textbf{gte-large} backbone, which left the main qualitative retrieval trends unchanged. We therefore retained all-MiniLM-L6-v2 for the reported experiments because of its lower computational cost. This preliminary comparison should not be interpreted as a systematic embedding-model ablation: a larger or domain-tuned model may change the similarity-score distribution, absolute retrieval quality, and the position of the detected elbow, even if the proposed adaptive-selection and graph-traversal mechanisms remain applicable. Long text is chunked at a stride of $240$ tokens, producing multiple \texttt{(vector, chunk\_text)} pairs per node. Let $\operatorname{chunks}(n)$ denote the set of text chunks associated with node $n$, $\mathbf{e}_q \in \mathbb{R}^{384}$ the embedding of query $q$, and $\mathbf{e}_c \in \mathbb{R}^{384}$ the embedding of chunk $c$. At retrieval time, the similarity of node $n$ to query $q$ is defined as the maximum cosine similarity across all chunks belonging to that node: \begin{equation} \operatorname{sim}(q,n) = \max_{c \in \operatorname{chunks}(n)} \frac{\mathbf{e}_q \cdot \mathbf{e}_c} {\|\mathbf{e}_q\|\,\|\mathbf{e}_c\|}. \label{eq:cosine-sim} \end{equation} 
Here, $\cdot$ and $\|\cdot\|$ denote the Euclidean inner product and norm. The maximum assigns each node the score of its best-matching chunk; nodes are then ranked by this score.

\subsection{Adaptive Elbow Detection}
\label{sec:elbow}

Equation~\eqref{eq:cosine-sim} scores every candidate node; sorting those scores in
descending order gives a similarity sequence $s_0 \geq s_1 \geq \cdots \geq s_{n-1}$, where
$n$ is the number of candidates in the pool and $s_i$ is the cosine similarity of the
candidate at rank $i$ (so $s_0$ is the closest match and $s_{n-1}$ the farthest). Such a
curve typically has a steep head of clearly relevant nodes followed by a long, flat
\emph{tail} of near-ties that are mostly noise; the ``elbow'' is the rank where the head
gives way to the tail. A fixed top-$k$ commits to that rank in advance, whereas we locate it
per question from the shape of the curve, in four steps.

\textbf{(1)~Smoothing.} Adjacent similarities are noisy---two candidates can swap places by
chance---which would jitter the detected elbow. We first pass a moving average over the
sequence. To avoid hand-tuning its width, we tie it to the pool size: a base window
$w_1 = \max(3,\lfloor\log_2 n\rfloor)$ and a wider one $w_2 = 2w_1 - 1$, both edge-padded so
the endpoints are not shortened. Smoothing at two widths, rather than one, keeps the outcome
from hinging on a single arbitrary choice.

\textbf{(2)~Elbow as the point of maximum bend.} For each smoothed sequence $\tilde{s}^{(k)}$, with $k\in\{1,2\}$ indexing the two windows, let $m_k$ denote its length. We draw the straight chord between its first and last points, $(0,\tilde{s}^{(k)}_0)$ and $(m_k-1,\tilde{s}^{(k)}_{m_k-1})$. The distance of the point at rank $i$ from this chord is \begin{equation} d_i^{(k)} = \frac{ \left| \left(\tilde{s}^{(k)}_{m_k-1}-\tilde{s}^{(k)}_0\right)i -(m_k-1)\tilde{s}^{(k)}_i +(m_k-1)\tilde{s}^{(k)}_0 \right| }{ \sqrt{ \left(\tilde{s}^{(k)}_{m_k-1}-\tilde{s}^{(k)}_0\right)^2 +(m_k-1)^2 } }, \qquad i\in\{0,\ldots,m_k-1\}. \label{eq:chord-distance} \end{equation} Here, $i$ is the rank and $\tilde{s}^{(k)}_i$ its smoothed similarity. The candidate elbow is the earliest rank attaining the maximum distance: \begin{equation} \hat{\imath}^{(k)} = \min\!\left\{ i : i\in \operatorname*{arg\,max}_{j\in\{0,\ldots,m_k-1\}} d_j^{(k)} \right\}. \label{eq:elbow-distance} \end{equation}

The detector uses chord distance rather than differential curvature. Its distance profile need not be unimodal and may contain multiple local maxima or tied global maxima. For a tie, we select the smallest index, avoiding an unnecessarily deep cutoff. Smoothing and aggregation across two windows reduce local instability but do not guarantee uniqueness; selection among distinct local maxima remains an open problem.
\textbf{(3)~Robustness via median.} The two windows can disagree on where the elbow sits; we reduce them to a single integer index \[ \hat{\imath} = \left\lfloor \operatorname{median}\!\bigl( \hat{\imath}^{(1)},\hat{\imath}^{(2)} \bigr) \right\rfloor. \] For two indices, the median is their midpoint, which is rounded down before applying the aggressiveness scaling. This aggregation prevents either smoothing width from determining the cutoff on its own.

\textbf{(4)~Aggressiveness scaling.} Steps (1)--(3) give a neutral, data-driven cut; a single
knob $\alpha>0$ then trades recall against precision without re-detecting anything, acting in
two reinforcing ways. It first shifts the elbow location according to \begin{equation} \hat{\imath}_\alpha = \operatorname{clip} \left( \left\lfloor \alpha\hat{\imath}\right\rfloor, 1, n-1 \right), \label{eq:aggressive-index} \end{equation} where $\alpha>0$ is the aggressiveness parameter and $\operatorname{clip}(x,a,b)=\min(\max(x,a),b)$. Values $\alpha>1$ move the cutoff deeper into the ranking and retain more candidates, whereas values $\alpha<1$ move it toward the head and retain fewer candidates. The lower bound of $1$ ensures that the highest-ranked candidate is always retained. Rather than cutting directly at the integer rank $\hat{\imath}_\alpha$, we convert it into the similarity threshold \begin{equation} \tau = \operatorname{clip} \left( s_{\hat{\imath}_\alpha} - (s_0-s_{\hat{\imath}_\alpha})(\alpha-1), s_{n-1}, s_0 \right). \label{eq:aggressive-threshold} \end{equation} The retriever retains candidates whose original similarity satisfies $s_i\geq\tau$. For $\alpha>1$, both transformations move the cutoff deeper and lower $\tau$; for $\alpha<1$, they have the opposite effect. At $\alpha=1$, the adjustment vanishes. When $n\leq1$, the available candidate is returned unchanged.

The elbow reads the shape of the similarity curve for the question at hand, so the
per-question budget it returns reflects how densely the corpus populates the knee, not the
number of gold tables (a point we return to in Section~\ref{sec:discussion}).

\subsection{Bidirectional Foreign-Key Traversal}
\label{sec:traversal}

After the initial table selection $\mathcal{T}$ via embedding similarity, the set is
expanded by traversing \texttt{REFERENCES} edges for $t_\text{hop}$ iterations. At each
hop, the algorithm adds all tables connected to the current frontier via
\texttt{REFERENCES} edges in either direction. The traversal is bidirectional because a
question can be asked from the perspective of either side of a one-to-many relationship.

The mechanism targets \emph{bridge tables}: tables required for a valid join despite weak semantic similarity to the question. In the scenario of Section~\ref{sec:graphstore}, \texttt{orders} and \texttt{order\_items} connect the semantically salient endpoint tables. Bidirectional traversal can recover them from either endpoint selected by the embedding stage.

\begin{algorithm}[H]
\caption{ElbowHop (Adaptive-Elbow Graph Traversal) Table Retrieval.}
\label{alg:graphrag}
\begin{algorithmic}[1]
\REQUIRE Query $q$, graph $G$, parameters $\alpha_\text{table}$, $t_\text{hop}$
\ENSURE Table set $\mathcal{T}$
\STATE $\mathbf{e}_q \leftarrow \textsc{Embed}(q)$
\STATE $\mathcal{T} \leftarrow \textsc{TableRAG}(G,\,\mathbf{e}_q,\,\alpha_\text{table})$ \COMMENT{cosine sim + adaptive elbow}
\FOR{$t = 1$ \TO $t_\text{hop}$}
  \STATE $\mathcal{T} \leftarrow \mathcal{T}\cup\textsc{ReferenceNeighbours}(G,\mathcal{T})$ \COMMENT{bidirectional FK traversal}
  \IF{no new tables added} \STATE \textbf{break} \ENDIF
\ENDFOR
\RETURN $\mathcal{T}$
\end{algorithmic}
\end{algorithm}

\section{Experimental Setup}
\label{sec:experiments}

\subsection{Corpora and the Pool-Size Axis}
\label{sec:corpora}

The central experimental axis is the size of the candidate pool the retriever must search.
We construct three corpora of growing pool size. \textbf{BIRD
per-DB} queries the BIRD Mini-Dev set~\cite{li2023bird} one database at a time, so each
question is answered against only its own $3$--$13$ tables; this is the easy regime in
which a fixed budget is already nearly correct. \textbf{BIRD merged} loads the same eleven
schemas into a single $74$-table property graph, so every question must now be
disambiguated against ten other databases. \textbf{Spider merged} loads the Spider
training databases~\cite{yu2018spider} into one $546$-table graph, the largest pool. The first transition (per-DB $\to$ merged) uses the same eleven BIRD schemas and the same query set, changing only whether each question is retrieved against its source database or against the combined $74$-table pool. It therefore provides our controlled comparison of pool size. Spider merged serves a different purpose: it tests whether the qualitative behaviour observed on BIRD persists on a second corpus and at a substantially larger scale. The transition from BIRD merged to Spider merged does not isolate pool size, since dataset composition changes as well, and should therefore not be interpreted as a controlled causal comparison of the $74$- and $546$-table conditions. Each corpus is evaluated on $500$ NL--SQL pairs (seed
$1234$). For each query the gold tables are extracted from the gold SQL with an abstract-syntax-tree
parser (\texttt{sqlglot}), falling back to a regular-expression parser only when AST
parsing fails. Every retriever consumes the same embeddings, graph store, and database
connection. 

\subsection{Metrics}
\label{sec:metrics}

Let $\hat{\mathcal{T}}$ denote the set of tables returned by a retriever and $\mathcal{T}^{\star}$ the set of gold tables extracted from the reference SQL query. We compute table-level recall and precision as \begin{equation} \operatorname{Recall} = \frac{ |\hat{\mathcal{T}}\cap\mathcal{T}^{\star}| }{ |\mathcal{T}^{\star}| }, \label{eq:recall} \end{equation} \begin{equation} \operatorname{Precision} = \frac{ |\hat{\mathcal{T}}\cap\mathcal{T}^{\star}| }{ |\hat{\mathcal{T}}| }. \label{eq:precision} \end{equation} Here, $|\cdot|$ denotes set cardinality and $\hat{\mathcal{T}}\cap\mathcal{T}^{\star}$ is the set of correctly retrieved gold tables.

Recall is the quantity a first-stage retriever must protect---a gold table dropped here is
generally unrecoverable downstream---while precision quantifies how much distractor budget the
generator (or a precision stage) must absorb, distractors having a measurable
cost~\cite{shi2023distracted,liu2024lostmiddle,cuconasu2024noise}.

\subsection{Retrievers and Ablation Variants}
\label{sec:retrievers}

All retrievers are backed by the same property-graph store and differ only in how they
select tables. To decompose the two mechanisms, we compare the full retriever against an
ablation of each. All baselines stay within the retriever's own cost class---local-embedder
methods that issue no LLM calls at retrieval time, the kind a production top-$k$ RAG
pipeline deploys; agentic and LLM-in-the-loop selectors occupy a different cost/latency
regime and are the downstream precision stage we assume, not a baseline here.

\begin{itemize}[leftmargin=*]
  \item \textbf{ElbowHop} (ours). Adaptive elbow with aggressiveness $\alpha_\text{table}$,
    followed by $t_\text{hop}$ iterations of bidirectional \texttt{REFERENCES} traversal
    (Algorithm~\ref{alg:graphrag}). The seed budget is set per question by the elbow.
  \item \textbf{ElbowHop-fixed$k$} (elbow ablation). Identical to ElbowHop but the
    per-question elbow seed is replaced by a fixed seed size equal to the mean elbow
    seed at that sweep cell, followed by the same FK traversal. Comparing ElbowHop against
    ElbowHop-fixed$k$ isolates the value of per-question adaptivity.
  \item \textbf{Dense (same-$k$)} (traversal ablation / baseline). Pure top-$k$ cosine
    ranking over the Table nodes, matched to the per-question total output of
    ElbowHop-fixed$k$: $k = |\mathcal{T}_\text{fixed-$k$}|$, the number of tables
    returned by ElbowHop-fixed$k$ after FK expansion (seed tables plus
    FK-expanded neighbours). Dense and ElbowHop-fixed$k$ therefore retrieve the same
    total number of tables per question; any recall difference is attributable solely to
    the FK structure, not to budget inflation.
  \item \textbf{Hybrid RRF}. Reciprocal rank fusion~\cite{cormack2009rrf} of the dense and
    a sparse Okapi BM25~\cite{robertson1995okapi} ranking
    ($k_1{=}1.5$, $b{=}0.75$) at the canonical $k_\text{rrf}{=}60$, budget-matched per
    question. BM25 tokenisation is lowercase $\texttt{[A-Za-z0-9\_]+}$, which splits
    snake\_case implicitly.
\end{itemize}

\subsection{Comparison Protocol and Parameter Sweep} \label{sec:protocol} Dense is matched per question to the total output of ElbowHop-fixed$k$, including FK expansion. Thus, the two methods return the same number of tables and their recall difference isolates the effect of graph traversal. In the full frontier comparison, ElbowHop retains its smaller adaptive budget by design. Reported $\langle k\rangle$ values are averages over questions. We sweep $\alpha_\text{table}\in\{0.2,0.4,0.6,0.8,1.0,1.2,1.4\}$ and $t_\text{hop}\in\{0,1,2,3\}$, yielding $14{,}000$ per-query records per corpus. Metrics are means over $500$ questions with $\pm1$ SEM bands. Unless stated otherwise, reported ablation values use $\alpha_\text{table}=0.6$. BM25 uses $k_1=1.5$ and $b=0.75$, while RRF uses $k_\text{rrf}=60$.


\section{Results}
\label{sec:results}

We first read the ElbowHop operating surface across the three pools
(Section~\ref{sec:res-overview}), then isolate each mechanism in turn: the adaptive elbow
versus a fixed budget (Section~\ref{sec:res-elbow}), the foreign-key traversal versus none
at a matched budget (Section~\ref{sec:res-hop}), and the full retriever against the
budget-matched baselines (Section~\ref{sec:res-combined}).

\subsection{The Operating Surface Across Pool Sizes}
\label{sec:res-overview}

Sweeping ElbowHop over the $t_\text{hop}\times\alpha_\text{table}$ grid, the qualitative
structure is identical across the three pools: table recall rises with both
$\alpha_\text{table}$ and $t_\text{hop}$ while precision falls. What changes dramatically
with pool size is the price of high recall. On BIRD per-DB, one hop already reaches
$\mathbf{81}$--$\mathbf{99}\%$ recall while precision stays at $\mathbf{34}$--$\mathbf{60}\%$; on BIRD merged the same one-hop
row spans $\mathbf{82}$--$\mathbf{100}\%$ recall but precision collapses from $\mathbf{57}\%$ to $\mathbf{3}\%$ as $\alpha$
grows; on Spider merged it reaches $\mathbf{100}\%$ recall only at precision near zero, with the
maximal budget approaching the entire $546$-table pool. The region of high recall at usable
precision is broad on the small pool and almost disappears on Spider.

We adopt $\alpha_\text{table}{=}0.6$ as the single default operating point for the analyses
that follow: at this setting table recall is already high on the realistic merged pools while
the set stays bounded (of order ten tables on BIRD merged, twenty on Spider merged). The
meaningful range $\alpha_\text{table}\!\in\![0.2,1.2]$ holds across all three corpora despite
pool sizes differing by two orders of magnitude; what changes across pools is the absolute
budget the elbow allocates at a given $\alpha$, while the useful range of $\alpha$ itself does
not.

\FloatBarrier
\subsection{The Adaptive Elbow Beats a Fixed Budget}
\label{sec:res-elbow}

\begin{figure}[!htbp]
  \centering
  \includegraphics[width=1\linewidth]{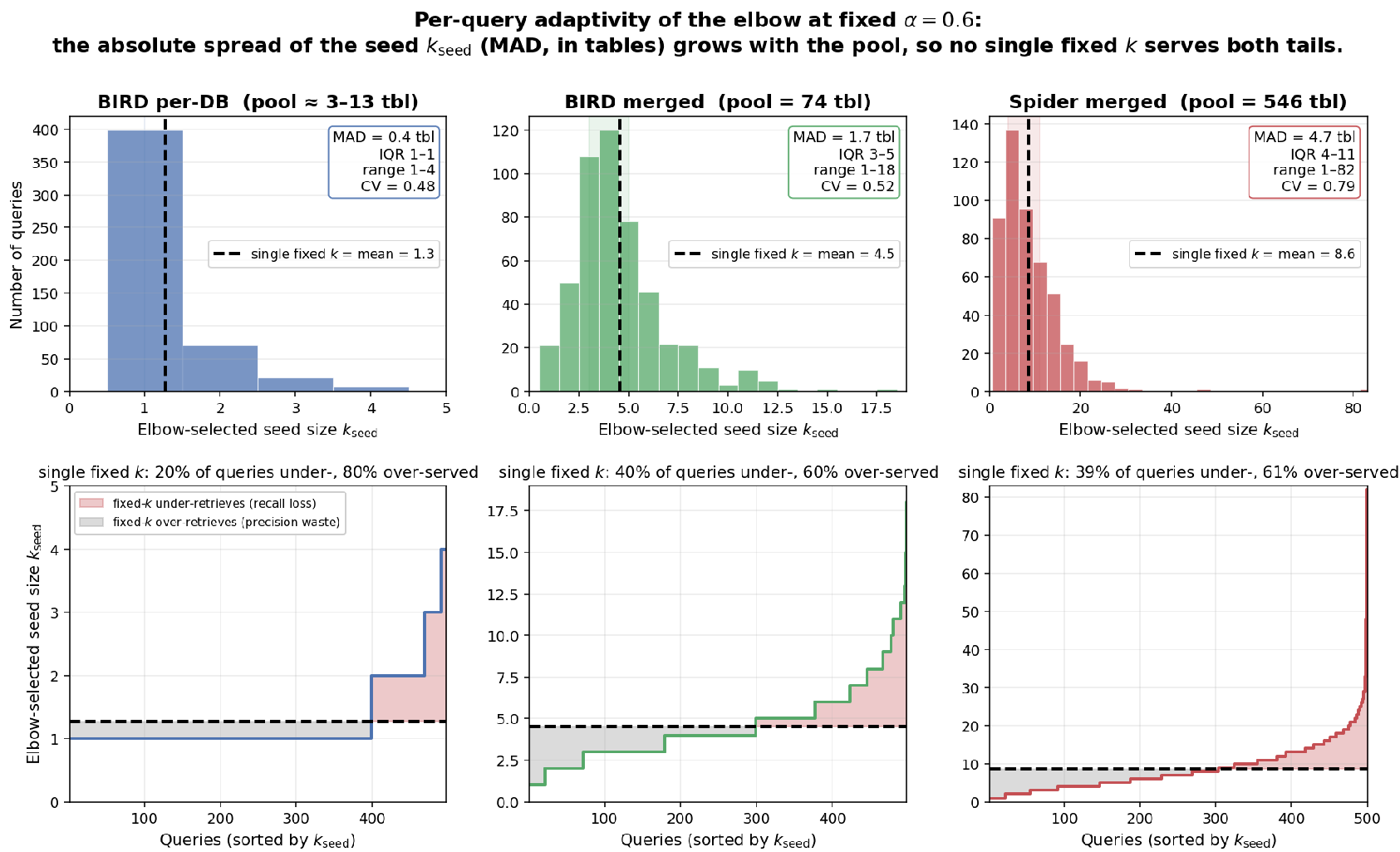}
  \caption{Per-question variation of the elbow seed at a single aggressiveness
  $\alpha_\text{table}{=}0.6$. \emph{Top:} distribution of the elbow seed $k_\text{seed}$ over
  $500$ questions per corpus; the dashed line is the mean (the ``fixed $k$''). The absolute
  spread (MAD, in tables) grows with the pool---$0.4\!\to\!1.7\!\to\!4.7$ tables---as does the
  relative spread (CV, $0.48\!\to\!0.52\!\to\!0.79$). \emph{Bottom:} questions sorted by
  $k_\text{seed}$; a single fixed $k$ (its mean) falls below the per-question seed for some
  queries (fewer tables, at risk to recall) and above it for the rest (more tables, at cost to
  precision), splitting them $20\%/80\%$ (BIRD per-DB), $40\%/60\%$ (BIRD merged), and
  $39\%/61\%$ (Spider merged).}
  \label{fig:elbow-var}
\end{figure}

Figure~\ref{fig:elbow-var} fixes $\alpha_\text{table}{=}0.6$ and plots the distribution of the
per-question elbow seed $k_\text{seed}$ over the $500$ questions. The elbow does not settle on
one value, and the right way to measure its spread is in absolute tables, since precision
is governed by $|\mathcal{T}^\star|/k$: a fixed $k$ that misbudgets a query by $\Delta k$ tables
wastes or forgoes $\Delta k$ tables. By that measure the spread grows sharply with the
pool---mean absolute deviation $0.4\!\to\!1.7\!\to\!4.7$ tables (coefficient of variation
$0.48\!\to\!0.52\!\to\!0.79$; seed range $1$--$4\!\to\!1$--$18\!\to\!1$--$82$)---so a fixed $k$
set to the mean disagrees with the per-question seed by ever more tables, falling below it for
some queries (at risk to recall) and above it for the rest (at cost to precision). This
establishes only that the room for a per-question policy grows with pool size; it does
not, on its own, show the elbow's choices are the right ones, since deviations are
measured against the elbow's own mean---reading them as ``errors'' would assume exactly what is
in question. Whether the variation is useful---whether it buys recall and precision
against the gold tables---is settled by the head-to-head below.

\begin{figure}[!htbp]
  \centering
  \begin{subfigure}{0.33\linewidth}
    \centering
    \includegraphics[width=\linewidth]{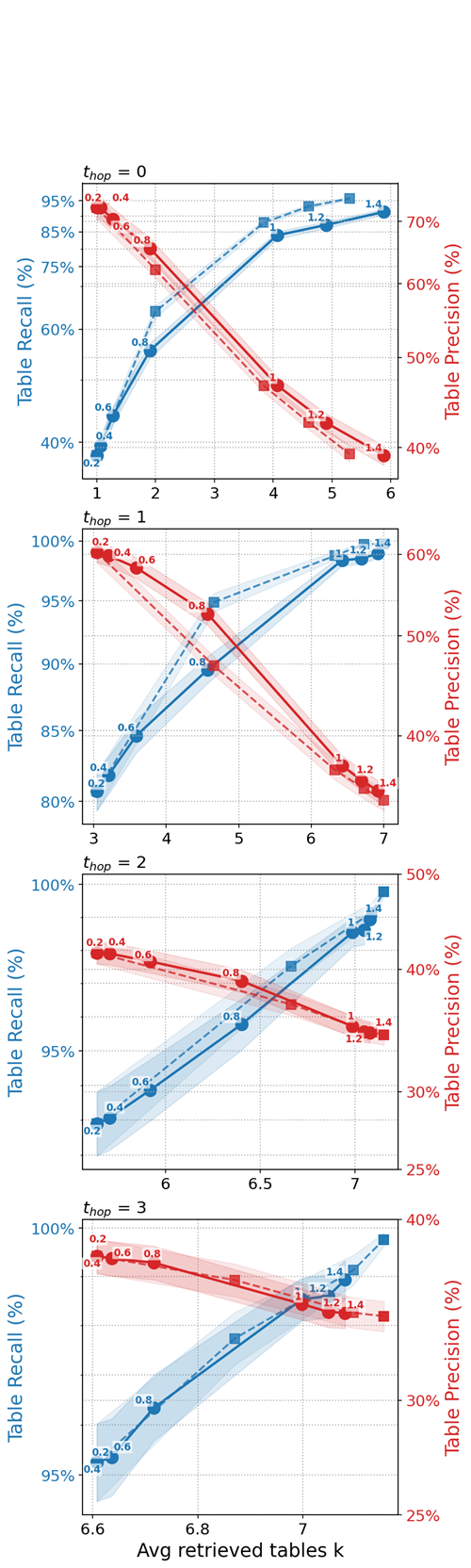}
    \caption{BIRD per-DB (3--13 tables).}
  \end{subfigure}\hfill
  \begin{subfigure}{0.33\linewidth}
    \centering
    \includegraphics[width=\linewidth]{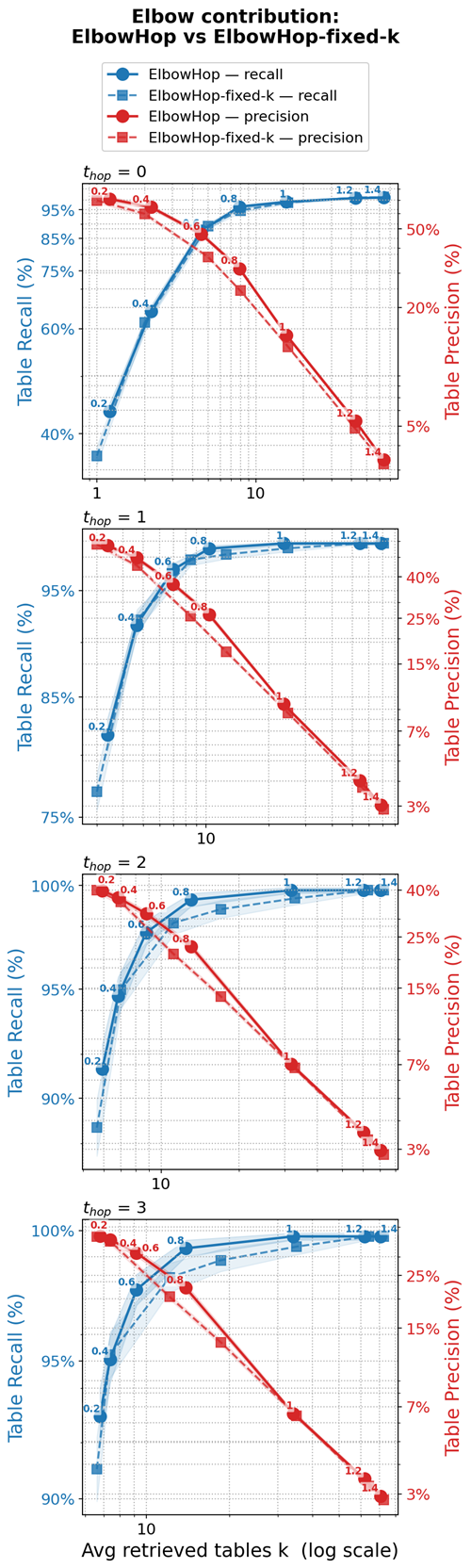}
    \caption{BIRD merged (74 tables).}
  \end{subfigure}\hfill
  \begin{subfigure}{0.33\linewidth}
    \centering
    \includegraphics[width=\linewidth]{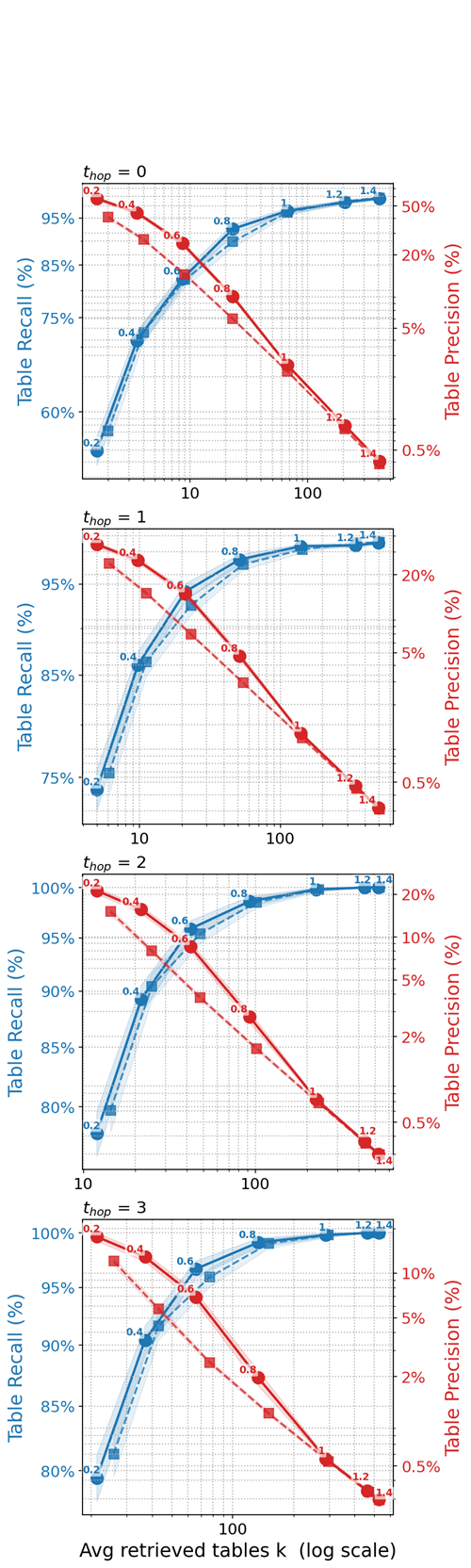}
    \caption{Spider merged (546 tables).}
  \end{subfigure}
  \caption{ElbowHop (per-question elbow, solid) versus
  ElbowHop-fixed$k$ (fixed seed, dashed), each at its own budget, with blue $=$ recall
  (left axis) and red $=$ precision (right axis), one panel per $t_\text{hop}$ and $\pm 1$
  SEM bands. At matched recall the elbow holds substantially higher precision, and the gap
  grows with pool size. Markers along each curve are the swept $\alpha_\text{table}$ values,
  labelled in every panel and increasing with $k$ from left to right ($\alpha_\text{table}{=}0.6$,
  our default, is the third).}
  \label{fig:elbow}
\end{figure}

Figure~\ref{fig:elbow} isolates the elbow by comparing ElbowHop (per-question elbow $+$ FK)
against ElbowHop-fixed$k$ (fixed seed $+$ FK), each at its own budget, one panel per hop depth
$t_\text{hop}\in\{0,1,2,3\}$. On the two merged pools, at essentially identical recall
the elbow delivers markedly higher precision at every hop depth: on BIRD merged at
$t_\text{hop}{=}1$ it reaches $\mathbf{97\%}$ recall at $\mathbf{36.9\%}$ precision versus $\mathbf{25.6\%}$ for the
fixed-$k$ seed, and the advantage widens with the pool---on Spider merged the elbow lifts
precision from $\mathbf{7.0\%}$ to $\mathbf{14.3\%}$ at one hop. On
the small BIRD per-DB pool the win does not reproduce: at $t_\text{hop}{\le}1$ the elbow's smaller
budget trades a little recall for precision rather than winning outright, and at
$t_\text{hop}{\ge}2$ the two budgets all but coincide and the curves converge. With only a handful
of candidate tables there is little room for a per-question budget to differ from a fixed one (the
$0.4$-table divergence of Figure~\ref{fig:elbow-var}), so the benefit is confined to the larger
pools and grows with their size.

\subsection{Foreign-Key Traversal Analysis}
\label{sec:res-hop}

\begin{figure}[!htbp]
  \centering
  \begin{subfigure}{0.33\linewidth}
    \centering
    \includegraphics[width=\linewidth]{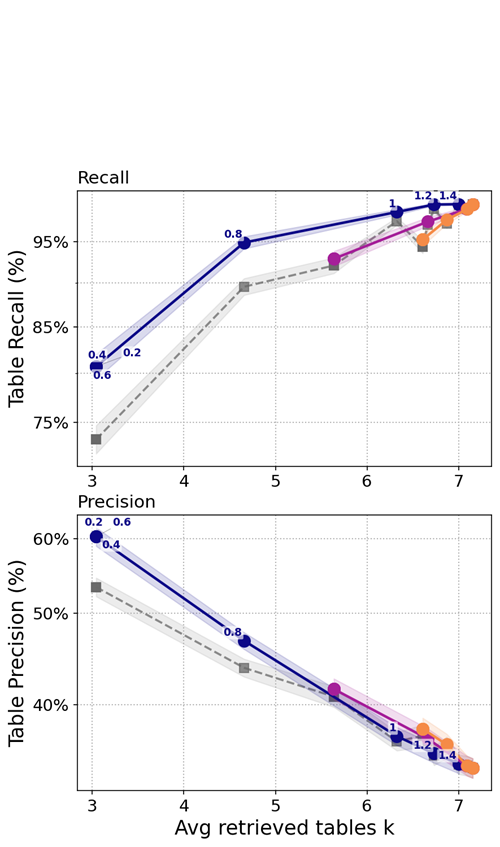}
    \caption{BIRD per-DB (3--13 tables).}
  \end{subfigure}\hfill
  \begin{subfigure}{0.33\linewidth}
    \centering
    \includegraphics[width=\linewidth]{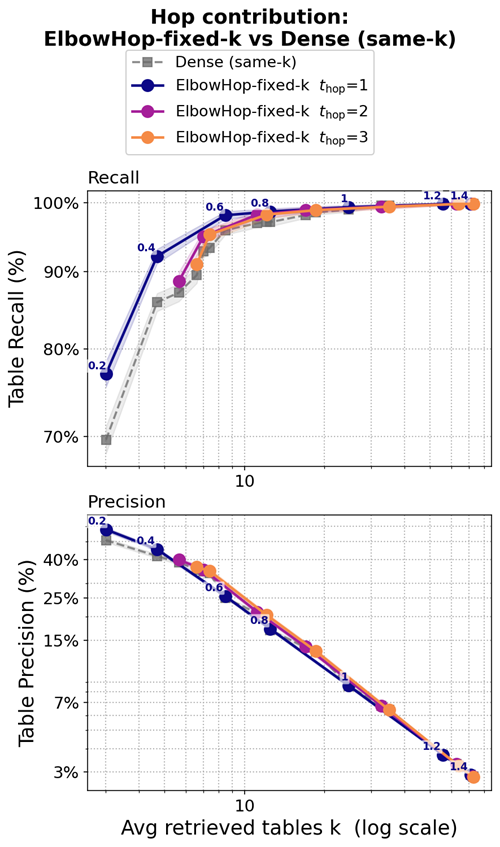}
    \caption{BIRD merged (74 tables).}
  \end{subfigure}\hfill
  \begin{subfigure}{0.33\linewidth}
    \centering
    \includegraphics[width=\linewidth]{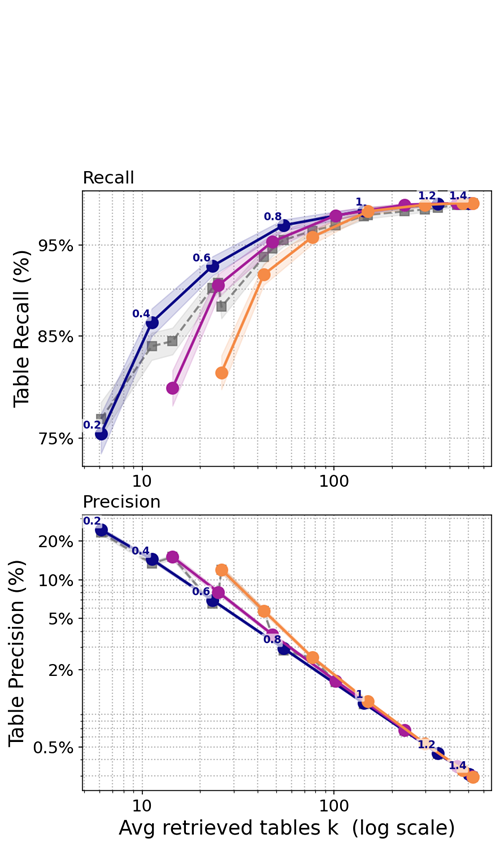}
    \caption{Spider merged (546 tables).}
  \end{subfigure}
  \caption{Foreign-key hop contribution at iso-budget. ElbowHop-fixed$k$ (seed $+$ FK, one
  curve per $t_\text{hop}$) versus Dense at the same per-question total budget
  ($k=|\mathcal{T}_\text{fixed-$k$}|$); recall (top) and precision (bottom) with $\pm 1$ SEM
  bands. The FK traversal adds only $+3$ to $+8$ recall points, at $t_\text{hop}{=}1$;
  precision is identical between the two.}
  \label{fig:hop}
\end{figure}

We isolate the traversal by comparing ElbowHop-fixed$k$ (fixed seed $+$ FK) against Dense
at the same per-question total budget: Dense retrieves $|\mathcal{T}_\text{fixed-$k$}|$
tables per question by pure cosine ranking---the same count as ElbowHop-fixed$k$ after FK
expansion. With the total table count identical, every recall difference is attributable to
the FK structure alone. Figure~\ref{fig:hop} plots the recall (top) and precision (bottom)
frontiers for all three pools. The recall gain from the FK structure is budget-dependent: it is largest at
the tight-budget end of each curve (low $\alpha_\text{table}$) and on the smaller
pools---where the seed is still impoverished, the FK-expanded set recovers more gold tables
than pure dense at the same total budget---and it narrows as the budget grows and the seed already
spans the join neighbourhood. 

At our $\alpha_\text{table}{=}0.6$ default the gap is already
modest ($\mathbf{+8}$ points on BIRD per-DB, $\mathbf{+2}$ on BIRD merged, $\mathbf{+3}$ on Spider merged), and a
single hop captures essentially all of it: the second hop adds $\mathbf{+1}$ or less on every corpus
($\mathbf{\approx 0}$ on Spider). On precision the two methods are indistinguishable, both lying
on the same $\approx |\mathcal{T}^\star|/k$ line (bottom panels). Aggregate recall averages
the join-critical bridge tables into the much larger set of content tables the dense seed
already finds, which we isolate next.

\begin{figure}[!htbp]
  \centering
  \begin{subfigure}{0.33\linewidth}
    \centering
    \includegraphics[width=\linewidth]{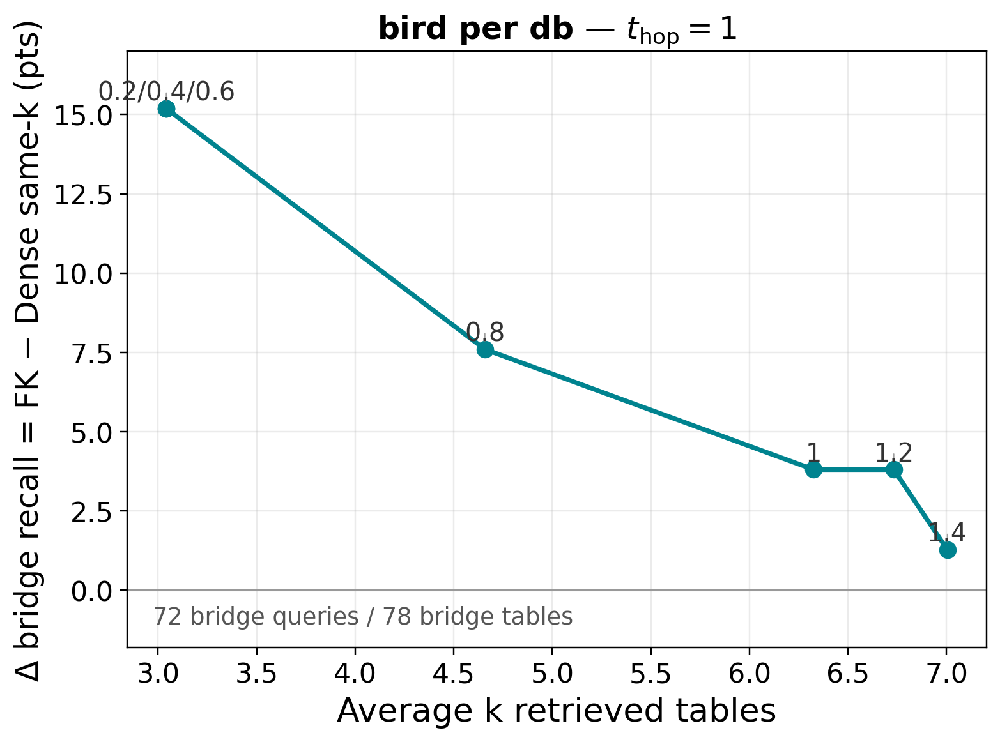}
    \caption{BIRD per-DB (3--13 tables).}
  \end{subfigure}\hfill
  \begin{subfigure}{0.33\linewidth}
    \centering
    \includegraphics[width=\linewidth]{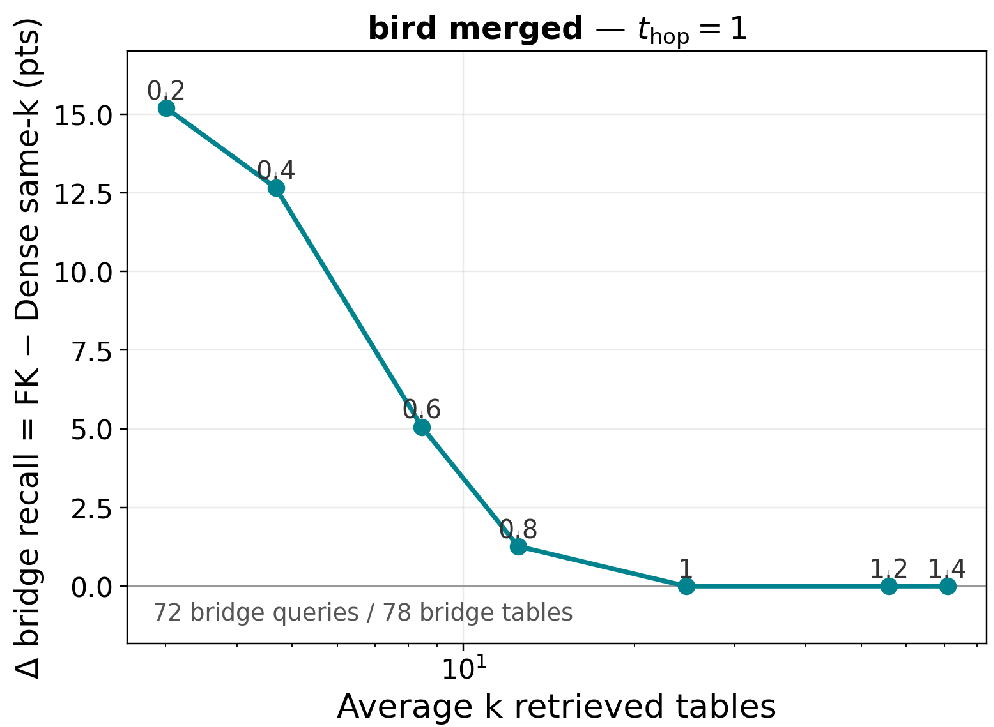}
    \caption{BIRD merged (74 tables).}
  \end{subfigure}\hfill
  \begin{subfigure}{0.33\linewidth}
    \centering
    \includegraphics[width=\linewidth]{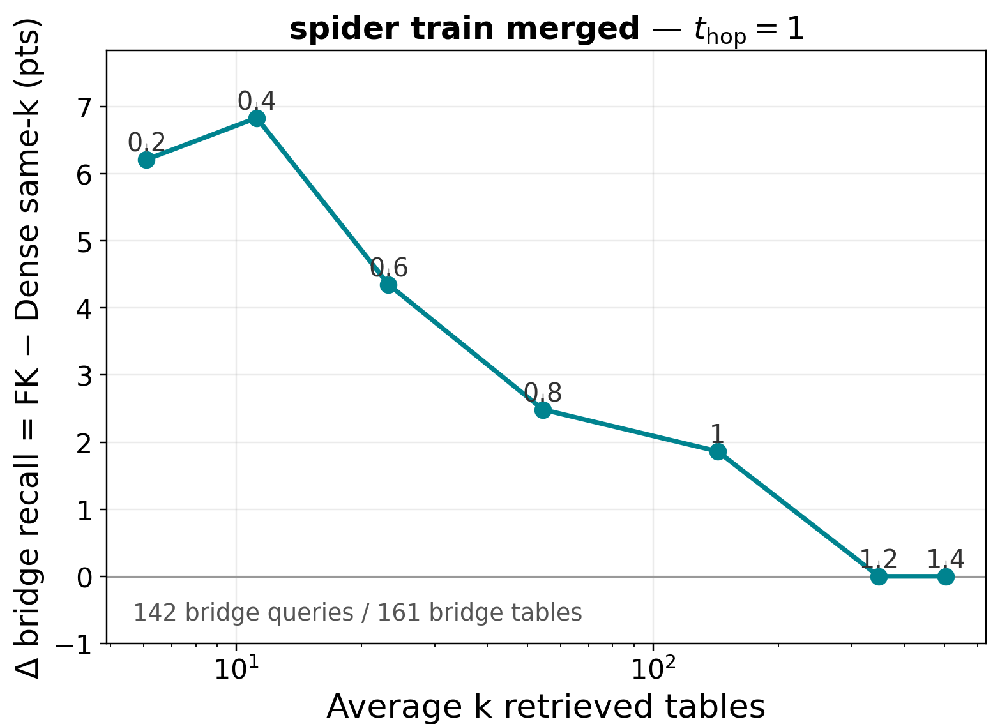}
    \caption{Spider merged (546 tables).}
  \end{subfigure}
  \caption{Bridge-table recall gain of the FK traversal at iso-budget:
  $\Delta=\text{recall}(\text{seed}{+}\text{FK})-\text{recall}(\text{dense same-}k)$, in points,
  restricted to content-column-free junction/linking gold tables, at $t_\text{hop}{=}1$. Each
  marker is one swept $\alpha_\text{table}$, placed at its mean budget $k$. The gain peaks at
  tight budget ($+15$ points on BIRD, $+6$--$7$ on Spider) and decays as $k$ grows. Counts:
  $72$ bridge queries / $78$ bridge tables on BIRD, $142$ / $161$ on Spider.}
  \label{fig:bridge}
\end{figure}

\paragraph{Where the traversal pays off: bridge tables.}
We re-analyse the saved iso-budget runs, restricting recall to
\emph{bridge tables}: gold tables that appear in the query yet contribute no content
column---none in \texttt{SELECT}, \texttt{WHERE}, \texttt{GROUP BY}, \texttt{HAVING}, or
\texttt{ORDER BY}---their only occurrences being join keys. These junction/linking tables
carry little or no direct lexical signal from the question for the dense encoder and are recovered by FK traversal by
construction; we identify them structurally and deterministically from each gold query.
Fixing $t_\text{hop}{=}1$ and sweeping $\alpha_\text{table}$, Figure~\ref{fig:bridge} plots
$\Delta=\text{recall}(\text{seed}{+}\text{FK})-\text{recall}(\text{dense same-}k)$ in points
against the mean budget. On every corpus the bridge-table gain is large and concentrated
at tight budget: $\mathbf{+15}$ points on BIRD (per-DB and merged) and $\mathbf{+6}$--$\mathbf{7}$ on Spider at the
smallest $\alpha_\text{table}$, decaying toward zero as the budget grows---once $k$ is wide
enough, dense recovers the bridges by pure spread.

\FloatBarrier
\subsection{The Full Retriever vs.\ Budget-Matched Baselines}
\label{sec:res-combined}

\begin{figure}[!htbp]
  \centering
  \begin{subfigure}{0.33\linewidth}
    \centering
    \includegraphics[width=0.95
    \linewidth]{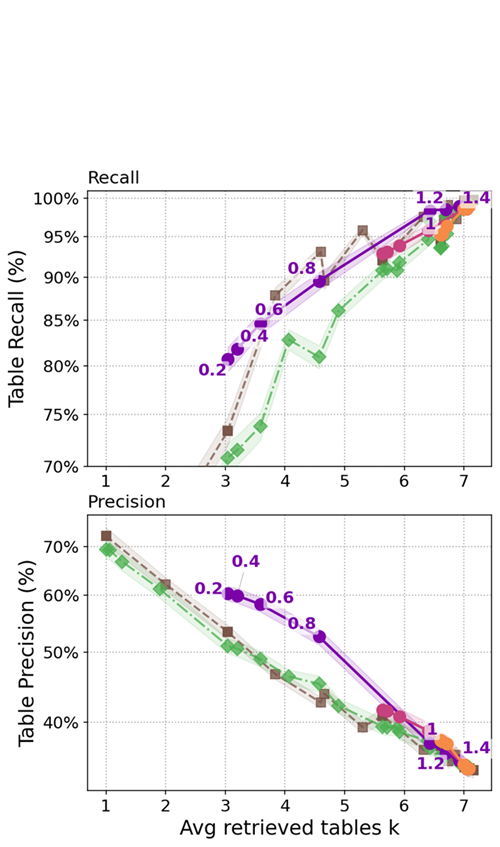}
    \caption{BIRD per-DB (3--13 tables).}
  \end{subfigure}\hfill
  \begin{subfigure}{0.33\linewidth} 
    \centering
    \includegraphics[width=0.95\linewidth]{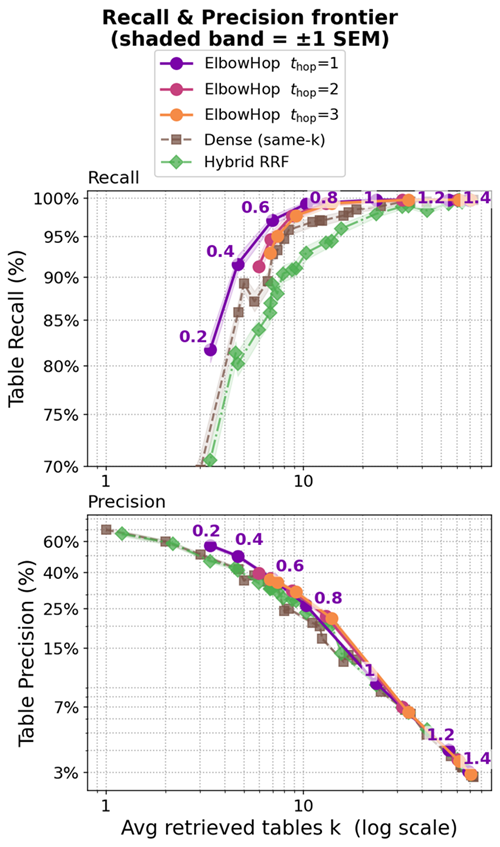}
    \caption{BIRD merged (74 tables).}
  \end{subfigure}\hfill
  \begin{subfigure}{0.33\linewidth}
    \centering
    \includegraphics[width=0.95\linewidth]{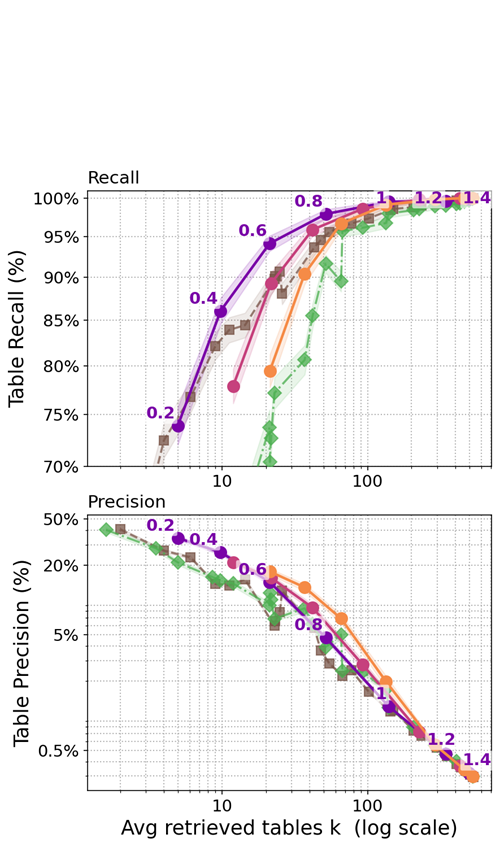}
    \caption{Spider merged (546 tables).}
  \end{subfigure}
  \caption{Recall (top) and precision (bottom) frontiers versus average budget $k$
  (log scale) for ElbowHop (solid, one per $t_\text{hop}$), Dense (dashed, at the
  ElbowHop-fixed$k$ total budget), and Hybrid-RRF (dashed), $\pm 1$ SEM bands. }
  \label{fig:frontier}
\end{figure}

Figure~\ref{fig:frontier} plots the recall and precision frontiers against the average
budget $k$ (log scale) for ElbowHop (one solid curve per $t_\text{hop}$) and the
Dense and Hybrid-RRF baselines. 

As described in Section~\ref{sec:protocol}, Dense sits at the higher fixed-$k$ budget and
ElbowHop at the lower elbow budget. Two facts hold across all three pools. First, at a given budget ElbowHop achieves higher recall than the baselines, thus reaching a target recall with fewer tables. This advantage widens with pool size, although the budget required for 99\% recall rises from $\langle k\rangle\approx10.4$ on BIRD merged to $\approx133.6$ on Spider merged. Second, at a fixed budget $k$ the methods trace a common precision envelope, so none is systematically more precise. ElbowHop's precision advantage appears at matched recall, because it reaches that recall with fewer tables. The collapse is tightest on the two BIRD pools; on Spider, ElbowHop curves fan out mildly by hop depth at matched $k$, while Dense and Hybrid stay on the same envelope.

Taking a target recall of $\geq 99\%$ as reference, Dense and Hybrid retrievers at their highest fixed-$k$ budget fall short of ElbowHop by roughly $\mathbf{1}$ to $\mathbf{6}$ recall points; the deficit is most pronounced on BIRD merged, where Hybrid lags by $\mathbf{6.2}$ points.
The precision there is whatever the budget
allows---about $\mathbf{35\%}$ on the small per-DB pool ($\langle k\rangle\!\approx\!6.9$), $\mathbf{26\%}$ on
BIRD merged ($\langle k\rangle\!\approx\!10.4$), and only $\mathbf{2\%}$ on Spider merged, where
reaching $99\%$ recall costs $\approx\!134$ tables.

\FloatBarrier

\section{Discussion}
\label{sec:discussion}

At the default operating point, the elbow provides most of the gain: it adapts the budget to the question-specific score profile, whose variability increases with pool size (Figure~\ref{fig:elbow-var}). Its useful $\alpha_\text{table}$ range remains stable across the evaluated pools, while the allocated budget changes with their scale. This conclusion is controlled for BIRD per-DB versus BIRD merged, which use the same schemas and questions; Spider merged instead provides a larger cross-corpus robustness check and does not isolate pool size from dataset composition. Additional matched-size subsets remain future work. The elbow does not estimate answer complexity: the correlation between $k_\text{seed}$ and the number of gold tables is negligible ($\rho\approx0.05$--$0.16$). FK traversal provides a smaller aggregate contribution but improves recall for semantically weak bridge tables, especially with one hop and a tight budget (Figure~\ref{fig:bridge}). At larger budgets, dense retrieval increasingly recovers the same tables. We therefore recommend a selective elbow with $t_\text{hop}=1$. At a fixed output size, precision is largely budget-bound. ElbowHop's advantage is therefore not higher precision at equal budget, but higher recall at equal budget, or equivalently, a smaller budget and higher precision at matched recall.

The present evaluation does not establish that this table-level advantage translates directly into higher SQL execution accuracy, since column selection and SQL generation are outside its scope. Nevertheless, at matched table recall, higher precision exposes the downstream generator to the same gold-table coverage with fewer irrelevant tables. Prior work shows that irrelevant or poorly positioned context can degrade model performance~\cite{shi2023distracted,liu2024lostmiddle,cuconasu2024noise}, which provides a plausible mechanism through which the smaller candidate sets produced by the elbow could reduce downstream generation errors. Smaller table sets also reduce the number of schema tokens passed to an LLM-based selector or generator. These remain expected downstream benefits rather than results demonstrated by the present experiments; an end-to-end evaluation of SQL validity and execution accuracy is left to future work.

Our current implementation completes the full retrieval pipeline in approximately 300--450\,ms per query. This measurement includes query embedding, table and column retrieval, elbow detection, in-memory FK traversal, and retrieval of stored SQL examples. By contrast, the 0.3--6\,ms measured for dense ranking covers only ranking over preloaded embeddings and metadata, with the query embedding already available. The two measurements therefore have different scopes and should not be interpreted as a controlled latency comparison or used to attribute the overhead to any individual component. The FK graph is loaded once from PostgreSQL and subsequently traversed in memory. A component-level latency breakdown is outside the scope of the present evaluation.

Another limitation is intrinsic to the elbow, which needs a sufficiently distinct knee to find. On short lists, or when similarities are uniformly high or low, the method may degrade toward returning most or all candidates, so extremely small or pathologically flat pools fall outside the regime where it helps. We did not separately label such flat-profile cases in the reported evaluation and therefore cannot provide a reliable estimate of their frequency. Potential diagnostic signals include a small similarity range $s_0-s_{n-1}$, a small maximum chord distance, or an elbow output approaching the full candidate pool. These quantities could support a fallback to a conservative fixed budget when the profile contains insufficient evidence for a distinct knee. However, defining and validating such a diagnostic threshold requires a dedicated analysis and is left to future work. Moreover, the chord-distance profile used by the detector is not guaranteed to have a unique maximum. Multiple local maxima may represent alternative transitions between the head and the tail of the ranking, and the global maximum is not necessarily the most operationally useful candidate if it occurs deep in the tail. The current method resolves tied global maxima conservatively by selecting the earliest one, but it does not explicitly compare distinct local maxima. Establishing regularity conditions under which the selected elbow is unique, and developing a principled criterion for choosing among multiple local maxima, remain directions for future work.
\section{Conclusion}
\label{sec:conclusion}

We presented ElbowHop, a table retriever combining an adaptive elbow cutoff with bidirectional FK traversal. Across pools of $3$--$13$, $74$, and $546$ tables, the elbow reaches matched recall with substantially higher precision than a fixed budget on the larger pools, while a single FK hop improves recovery of bridge tables at tight budgets. The method is best used as a recall-oriented first stage before downstream pruning or generation. Future work includes end-to-end SQL evaluation, flat-profile fallback diagnostics, and selection among multiple elbow candidates.
 
\begin{acknowledgments}
  This work is framed within the Project ``AVANT'' -- Project no. IPCEI-CL\_0000005 -- Application protocol no. 108421 of 14/05/2024 -- CUP B89J24002920005 -- Grant decree no. 1322 of August 8, 2024 -- financed by the European Union -- NextGenerationEU (IPCEI Funding).
\end{acknowledgments}

\section*{Declaration on Generative AI}
During the preparation of this work, the authors used a large language model assistant for grammar, spelling, and rephrasing, then reviewed and edited the content and take full responsibility for it.

\end{document}